\documentclass[superscriptaddress, amsmath,amssymb,aps,pra,twocolumn]{revtex4-1}

\usepackage{graphicx}
\usepackage{dcolumn}
\usepackage{bm}
\usepackage{xcolor}
\usepackage{amsmath,amsfonts,amsthm,bm}

\begin{document}

\preprint{APS/123-QED}

\title{Quantum nondemolition measurements in the relativistic Dirac oscillator}

\author{Wei Sun}
\affiliation{Quantum Institute for Light and Atoms, State Key Laboratory of Precision Spectroscopy, School of Physics and Electronic Science, East China Normal University, Shanghai 200241, China}

\author{Keye Zhang}
\email{kyzhang@phy.ecnu.edu.cn}
\affiliation{Quantum Institute for Light and Atoms, State Key Laboratory of Precision Spectroscopy, School of Physics and Electronic Science, East China Normal University, Shanghai 200241, China}

\author{Pierre Meystre}
\affiliation{J. Wyant College of Optical Sciences and Department of Physics, University of Arizona, Tucson, Arizona 85721, USA.}

\author{Weiping Zhang}
\email{wpz@sjtu.edu.cn}
\affiliation{School of Physics and Astronomy, and Tsung-Dao Lee Institute, Shanghai Jiao Tong University, Shanghai 200240, China}
\affiliation{Collaborative Innovation Center of Extreme Optics, Shanxi University, Taiyuan, Shanxi 030006, China}
\affiliation{Shanghai Research Center for Quantum Sciences, Shanghai 201315, China}

\date{\today}

\begin{abstract}
We investigate the feasibility of performing quantum nondemolition (QND) measurements in the one-dimensional Dirac oscillator,  a frequently used toy model to investigate relativistic extensions of nonrelativistic quantum effects. We derive general expressions for its QND observables and find that they are in general complex combinations of the position, momentum, and spin operators, which makes them challenging to realize experimentally. However, the situation is considerably simplified in both the weakly and strongly relativistic limits, where they should be amenable to an experimental demonstration. This suggests
exploiting continuous measurements of small departures from the QND character of these variables in the intermediate regime as probes of relativistic effects.
\end{abstract}

\maketitle

\section{Introduction}

Quantum nondemolition (QND) measurements were first proposed by Braginsky~\cite{braginsky1967classical,braginsky1969quantum} as a way to circumvent the Standard Quantum Limit of measurements, that is, the backaction on a second measurement of a quantum observable $\hat a(t)$ from the uncertainty imposed on its conjugate variable. Under normal circumstances the uncertainty in a second measurement of an observable $\hat A$ is limited by the Heisenberg Uncertainty Principle. QND variables are variables that commute with themselves at the times of $t_0$ and $t_1$ of these measurements, $[\hat A(t_0),\hat A(t_1)]=0$.  If this condition is satisfied at all times, $d \hat A /dt=0$, $\hat A(t)$  is called a continuous QND observable, else it is a stroboscopic QND observable see e.g. Refs.~\cite{braginsky1967classical,braginsky1969quantum,braginsky1975unperturbed,braginsky1980quantum,PhysRevLett.40.667}.

Reference.~\cite{PhysRevLett.40.667}, see also the reviews \cite{RevModPhys.52.341} and \cite{RevModPhys.68.755}, showed that in the case of a simple harmonic oscillator of mass $m$,  frequency $\omega$ and Hamiltonian $\hat H_0 = \hbar \omega \hat a^\dagger \hat a$, either of the two quadratures
\begin{eqnarray}
\hat Y_1 &=&\hat x \cos \omega t - (\hat p/m\omega)\sin \omega t, \nonumber \\
\hat Y_2 &=&\hat x \sin \omega t + (\hat p/m\omega)\cos \omega t, \label{X12} \, 
\label{eq:quadrat}
\end{eqnarray}
with $[\hat Y_1,\hat Y_2]=i\hbar/m\omega$,  $\Delta Y_1\Delta Y_2\geq \hbar/2m\omega$, and $d\hat Y_\ell/dt =0$, can be chosen as a QND observable (but of course not both simultaneously). This property was  exploited in particular in optical and atomic interferometry~\cite{chen2017quantum}, weak force and gravitational wave detection~\cite{chen2011qnd, Suh1262}, and the preparation of squeezed states~\cite{vasilakis2015generation}.

The goal of this paper is to extend these considerations to the case of a Dirac oscillator, an extension of the free particle Dirac equation obtained by the addition of a term linear in position and governed by the  equation
\begin{equation}
i \hbar \partial_t \psi = \left [ c {\bm \alpha} \cdot \left (  {\bf p}- i m \omega{\bf r} \beta \right ) + \beta mc^2 \right ] \psi.
\end{equation}
Here ${\bf p} = -i \hbar {\bm \nabla}$ and 
\begin{equation}
\bm \alpha = 
\begin{pmatrix}
0 &  {\bm \sigma} \\
{\bm \sigma} & 0 \\
\end{pmatrix} 
\,\,\,\,\,;\,\,\,\,\,
\beta = 
\begin{pmatrix}
I &  0 \\
0 & -I \\
\end{pmatrix}
\end{equation} 
with $\bm \sigma$ the vector Pauli spin matrix, so that $\alpha_i$  and $\beta$ satisfy the anticommutation relations ($i\neq j = x,y,z$)
\begin{equation}
\begin{aligned}
\alpha_{i}\alpha_{j} &=-\alpha_{j}\alpha_{i}\, ,\\
\alpha_{i}\beta&=-\beta\alpha_{i} \, ,
\end{aligned}
\end{equation}
and $\alpha_i^2 =\beta^2=1$.

The Dirac oscillator is characterized by the presence of a spin-orbit coupling term, as becomes particularly apparent when expressing the Dirac wave function in terms of its large and small components $\psi_1$ and $\psi_2$ as~\cite{Moshinsky_1989}
\begin{equation}
\psi = 
\begin{pmatrix}
\psi_1 \\
\psi_2 \\
\end{pmatrix}
 e^{-iEt/\hbar}
\end{equation}
with
\begin{eqnarray}
(E-mc^2)\psi_1 &=& c {\bm \sigma} \cdot({\bf p} + i m \omega {\bf r}) \psi_2 \nonumber \\
(E+mc^2)\psi_2 &=& c {\bm \sigma} \cdot({\bf p} - i m \omega {\bf r}) \psi_1 \,
\end{eqnarray}
exhibiting the familiar spin-orbit coupling of the Dirac equation. Reexpressing the energy $E$ as $E = mc^2 + {\cal E}$, one finds that in the weakly relativistic limit ${\cal E} \ll mc^2$  and when expressed in terms of the large component $\psi_1$ only, the energy of the system takes the form
\begin{equation}
{\cal E} =  \frac{\bf p^2}{2m} + \frac{m\omega^2 \bf r^2}{2} - \frac{2 \omega}{\hbar} {\bf L}\cdot {\bf S} -\frac{3\hbar \omega}{2},
\label{eq:LS}
\end{equation}
where ${\bf L} = {\bf r} \times {\bf p}$ and ${\bf S} = (\hbar/2) {\bm \sigma}$~\cite{Moshinsky_1989}. That is, in addition to the familiar energy of a simple harmonic oscillator, $\cal E$ comprises a spin-orbit coupling component of strength $-2 \omega/\hbar$. It is this property that justifies calling this system a Dirac oscillator. 

The Dirac oscillator is frequently used as a toy model to investigate relativistic extensions of nonrelativistic quantum effects. These include in particular the generation of Dirac cat states~\cite{PhysRevLett.99.123602}, the study of a relativistically deformed uncertainty principle in (1+1) dimensions~\cite{benzair2012path}, and a relativistic quantum heat engine using a single Dirac particle trapped in an infinite one-dimensional potential as its working substance~\cite{PhysRevE.86.061108}. It has also found wide applications in nuclear physics, condensed matter physics and quantum optics, see e.g. the review~\cite{quesne2017dirac}. 

The spin-orbit coupling characteristic of the Dirac equation is also central to on-going efforts to control the spin and electronic ground state topology in quantum systems, which requires its detailed characterization \cite{PhysRevLett.98.266801}, and is also becoming an efficient mechanism of quantum sensing in hybrid optomechanical systems \cite{PhysRevLett.119.233602,Hee1600485}.  For this reason, extending the familiar physics associated with QND measurements in the simple harmonic oscillator to this system can help clarify the difficulties associated with this additional interaction and suggest ways to exploit this complication in diagnostic tools.

Because of its nonequidistant excitation spectrum and the presence of spin-orbit coupling, the dynamics of the Dirac oscillator differs significantly from that of the simple harmonic oscillator. For example, we showed in earlier work~\cite{PhysRevLett.121.110401} that the interference between positive and negative energy states prevents the realization of quantum-mechanics free subsystems that avoid quantum measurement backaction. But whether a more direct generation of QND observables is possible remains an open question.

\section{Model}

We consider for simplicity the (1+1)-dimensional version of the Dirac oscillator.  In that case the Dirac operators $\bm \alpha$ and $\beta$, reduce to $2 \times 2$ matrices that can be taken as any two of the three Pauli matrices, for example,
\begin{equation}
 \alpha_{x} \to \hat\sigma_{x}\;\;;\;\; \beta \to \hat\sigma_{z}.\label{alphax} 
\end{equation}
and the Dirac oscillator is governed by the equation
\begin{equation}
i \hbar \frac{\partial \psi}{\partial t}=\left[ c\hat{\sigma}_x\hat{p}-mc\omega\hat{\sigma}_y\hat{x}+mc^2\hat \sigma_z \right ]\psi \equiv \hat H_{\rm D} \psi .
\label{1+1 D-equation}
\end{equation}

We introduce the bosonic annihilation and creation operators $\hat a$ and $\hat a^\dagger$ and the excitation number operator $\hat n=\hat a^\dagger \hat a$ in the usual way via $\hat x=x_{\rm zpt}(\hat a+\hat a^\dagger)$ and $\hat p=-ip_{\rm zpt}(\hat a-\hat a^\dagger)$, where  $x_{\text {zpt}}=\sqrt{\hbar/2m\omega}$ and  $p_{\text {zpt}}=\sqrt{m\hbar\omega/2}$ are the zero-point position and momentum. $\hat H_D$  takes then the form of the anti-Jaynes-Cumming Hamiltonian
\begin{equation}
\hat{H}_{\text D}=\eta\hat{\sigma}_{-}\hat{a}+\eta^\ast\hat{\sigma}_{+}\hat{a}^\dagger +mc^2\hat{\sigma}_{z},
\label{eq:antiJC}
\end{equation}
where
\begin{equation}
\eta=-imc^2 \sqrt{2\epsilon}\;\;\;\;;\;\;\;\;\epsilon = \hbar \omega/mc^2 .
\end{equation} 

As in (3+1) dimensions, the oscillator wave functions $\psi$ are spinors with components $\psi_1$ and $\psi_2$. With
\begin{equation}
\psi = 
\begin{pmatrix}
\psi_1 \\
\psi_2 \\
\end{pmatrix}
 e^{-iEt/\hbar},
\end{equation}
Eq.~(\ref{1+1 D-equation}) takes the form of the coupled equations
\begin{equation}
\begin{aligned}
(E-mc^2)\psi_{1}&=c(\hat{p}+im\omega\hat{x})\psi_{2},\\
(E+mc^2)\psi_{2}&=c(\hat{p}-im\omega\hat{x})\psi_{1}.
\end{aligned}
\end{equation}
or, in terms of $\hat a$ and $\hat a^\dagger$,
\begin{equation}
\begin{aligned}
(E-mc^2)\psi_{1}&=\eta^* \hat a^\dagger \psi_{2},\\
(E+mc^2)\psi_{2}&=  \eta \hat a\, \psi_{1}, 
\end{aligned}
\end{equation}
which explicitly show that the frequency $\omega$ is a measure of the spin-orbit coupling between $\psi_1$ and $\psi_2$ already encountered in Eq.~(\ref{eq:LS}).

The eigenstates of the Hamiltonian~(\ref{eq:antiJC}) are the dressed states 
\begin{eqnarray}
|E_n^+\rangle&=&A_{n}|n,\uparrow\rangle-iB_{n}|n-1,\downarrow\rangle ,\nonumber \\
|E_n^-\rangle&=&B_{n+1}|n+1,\uparrow\rangle+iA_{n+1}|n,\downarrow\rangle \, , \label{dressedstate}
\end{eqnarray} 
with associated positive and negative eigenenergies 
\begin{equation}
E_n^-=-E_{n+1}^+=-mc^2\sqrt{1+2(n+1)\epsilon} \, .
\end{equation}
Here $\left|\uparrow \right\rangle,\, \left| \downarrow \right\rangle$ are Pauli spinors, the states $\left|n\right\rangle$ are eigenstates of the nonrelativistic harmonic oscillator, and 
\begin{eqnarray}
A_{n}&=&\sqrt{(E_{n}^{+}+mc^2)/2E_{n}^{+}}\nonumber,\\
B_{n}&=&\sqrt{(E_{n}^{+}-mc^2)/2E_{n}^{+}} .
\label{eq:AnBn}
\end{eqnarray}
The asymmetry between the positive and negative energy branches and the property that the eigenstates are linear superpositions if motional states of opposite spins are direct consequences of the relativistic spin-coupling component of $\hat H_{\rm D}$.

\section{QND observables}
This section derives the general form of a pair of QND variables of the Hamiltonian $\hat H_{\rm D}$ and determines their dependence on the relativistic level of the oscillator, as characterized by the parameter 
\begin{equation}
\epsilon = \hbar \omega/mc^2,
\end{equation}
 with $\omega$ the oscillator frequency and $m$ its rest mass. We consider first the two limiting cases of the weakly relativistic regime, $\epsilon \ll 1$, and the extreme relativistic regime, $\epsilon \gg 1$, and show that the corresponding QND variables reduce then to the familiar quadrature operators $\hat Y_{1,2}$ of the simple harmonic oscillator, albeit with spin-dependent frequencies. In the intermediate regime, however, they take the form of intricate spin-orbit coupling operators. The following section will give estimates of experimental parameters achievable in QND measurements of implementations of the Dirac oscillator with either electrons or cold atoms.

\subsection{Weakly relativistic limit}
In the weakly relativistic limit $\epsilon \ll 1$ we have $E_n^+\approx mc^2+n\hbar\omega$, $A_n \approx 1$ and $B_n \approx 0$, so that the superpositions of opposite spin states disappear, the eigenstates reducing to $\left|n,\uparrow\right\rangle$ and $i\left|n, \downarrow\right\rangle$, respectively. This can be seen more explicitly when eliminating $\psi_2$ or $\psi_1$ from these two equations by multiplying them by $(E+mc^2)$ and $(E-mc^2)$, respectively, resulting in the Klein-Gordon-like equations 
\begin{equation}
\begin{aligned}
(E^2-m^2c^4)\psi_{1}&=2mc^2\left (\frac{\hat{p}^{2}}{2m}+\frac{1}{2}m\omega^2\hat{x}-\frac{\hbar\omega}{2}\right )\psi_{1},\\
(E^2-m^2c^4)\psi_{2}&=2mc^2\left (\frac{\hat{p}^{2}}{2m}+\frac{1}{2}m\omega^2\hat{x}+\frac{\hbar\omega}{2}\right )\psi_{2}.
\end{aligned}
\label{eq:squared}
\end{equation}
Consider first the component $\psi_1$ of the Dirac spinor. With  $E=mc^2+{\cal E}$, the weakly relativistic regime corresponds to ${\cal E} \ll mc^2$, so that
\begin{equation}
(E^2-m^2c^4)\approx 2mc^2 {\cal E}+O({\cal E}^2) 
\end{equation}
so that 
\begin{equation}
{\cal E}\psi_{1}=\left (\frac{\hat{p}^{2}}{2m}+\frac{1}{2}m\omega^2\hat{x}-\frac{\hbar\omega}{2}\right )\psi_{1}.
\end{equation}
with corresponding effective positive energy Hamiltonian
\begin{equation}
\hat{H}_{\rm{eff}}^{+}=mc^2+\left (\frac{\hat{p}^{2}}{2m}+\frac{1}{2}m\omega^2\hat{x}-\frac{\hbar\omega}{2}\right).
\label{1+1 up D-eq}
\end{equation}
Similarly, for the component $\psi_{2}$ we set $E=-mc^2+{\cal E}$. The weakly relativistic limit $|{\cal E} |\ll mc^2$ yields then
\begin{equation}
(E^2-m^2c^4)\approx -2mc^2 {\cal E}+O({\cal E}^2),
\end{equation}
so that
\begin{equation}
{\cal E}\psi_{2}=-\left (\frac{\hat{p}^{2}}{2m}+\frac{1}{2}m\omega^2\hat{x}+\frac{\hbar\omega}{2}\right )\psi_{2} 
\end{equation}
with corresponding negative energy Hamiltonian
\begin{equation}
\hat{H}_{\rm{eff}}^{-}=-mc^2-\left (\frac{\hat{p}^{2}}{2m}+\frac{1}{2}m\omega^2\hat{x}+\frac{\hbar\omega}{2}\right ).
\label{1+1 down D-eq}
\end{equation}
Combining Eqs.~(\ref{1+1 up D-eq}) and (\ref{1+1 down D-eq}) gives finally the effective Hamiltonian characterizing the weakly relativistic limit of the (1+1)-dimensional Dirac oscillator as
\begin{eqnarray}
\hat{H}_{\rm nr}&=&\left [mc^2+\frac{\hat{p}^{2}}{2m}+\frac{1}{2}m\omega^2\hat{x} \right ] \hat{\sigma}_z - \frac12 \hbar\omega \nonumber \\
&=& mc^2\hat\sigma_z+\hbar \omega \left (\hat a^\dagger \hat a + \frac12\right )\hat{\sigma}_z -\frac12 \hbar\omega .
\label{eq:Hnr}
\end{eqnarray}
In that limit the two components of the spinor $\psi$ become therefore approximately decoupled and $\hat \sigma_z$ is a constant of motion, hence a QND observable. In addition, it also follows that the system possesses an additional pair of QND observables $\hat{X}_{\ell, \text{nr}}$,  $\ell = (1,2)$, of precisely the same form as the quadrature operators $\hat Y_{\ell}$ of Eqs.~(\ref{X12}) but with the important difference that the frequency $\omega$ becomes now a spin-dependent frequency operator 
\begin{equation}
\omega \to \hat \omega_{\rm{nr}} \equiv \omega\hat\sigma_z,
\end{equation}
so that
\begin{eqnarray} 
\hat{X}_{1, \text{nr}}&=&x_{\text {zpt}}(\hat{a}e^{i\hat\omega_{\text{nr}}t}+ {\rm h.c.})\, ,\nonumber \\
\hat{X}_{2, \text{nr}}&=&i x_{\text{zpt}}(e^{-i\hat\omega_{\text{nr}}t}\hat{a}^\dagger  - {\rm h.c.}).
\label{eq:Xnr}
\end{eqnarray}
Since $[\hat a, \hat \omega_{\text{nr}}]=0$ these operators obey the same commutation relations as the harmonic oscillator quadratures $\hat Y_{1,2}$.

\subsection{Extreme relativistic limit}

In the other extreme limit, $\epsilon\gg 1$, the spin-orbit coupling term $\eta\hat{\sigma}_{-}\hat{a}+\eta^\ast\hat{\sigma}_{+}\hat{a}^\dagger$ dominates the Hamiltonian so that $E_n^+\approx mc^2\sqrt{2n\epsilon}$ and  $A_n\approx B_n\approx 1/\sqrt{2}$. Moreover, the energy difference between neighboring motional states $\left|n\right\rangle$ and $\left|n-1\right\rangle$ scales as $1/\sqrt{n}$ and approaches zero for large excitations $n\gg 1$. The eigenstates of $\hat H_{\rm D}$ can then be approximated by the factorized motional and spin states $\left | E_n^\pm\right\rangle\approx |n\rangle(\left |\uparrow\right\rangle\mp i\left |\downarrow\right\rangle)/\sqrt{2}$, corresponding to the effective Hamiltonian
\begin{equation}
\hat{H}_{\rm r}=-\hat{\sigma}_{y}\sqrt{2mc^2\hbar\omega\hat{n}} \, .
\label{eq:Hr}
\end{equation}
In contrast to $\hat H_{\rm nr}$ the eigenenergy spectrum of $\hat{H}_{\rm r}$ are not equidistant, and $\hat \sigma_y$ rather than $\hat \sigma_z$  is now a constant of motion, hence a QND observable. Like in the weakly relativistic case $\hat{H}_{\rm r}$ also possesses a pair of QND observables $\hat X_{\ell, \rm r}$ that are formally identical with the quadrature operators (\ref{eq:Xnr}), but now with the replacement of $\omega$ by a frequency operator $\hat\omega_{\rm r}$ with both a motional and spin dependence,
\begin{equation}
\omega \to \hat{\omega}_{\rm r} \equiv \omega\sqrt{2/\varepsilon}(\sqrt{\hat{n}-1}-\sqrt{\hat{n}})\hat{\sigma}_{y}\, .
\label{eq:omegar}
\end{equation}
With the equality $\hat{a}f(\hat{n})=f(\hat{n}+1)\hat{a}$ one can readily show that the dynamical evolution and uncertainty relation of $\hat X_{\ell, \rm r}$ are the same as those of the quadrature operators $\hat Y_\ell$. (Note that since $[\hat a, \hat \omega_{\text r}]\neq 0$ the operator ordering in $\hat X_{\ell, \rm r}$ cannot be arbitrarily changed.)

\subsection{General case}
It is easily verified that the formal similarity between the QND observables $\hat X_\ell$ of the Dirac oscillator in the limits $\epsilon \ll 1$ and $\epsilon \gg 1$ and the quadratures $\hat Y_\ell$ of the simple harmonic oscillator follows from the fact that in both limits $\hat n$ is a conserved quantity (approximately so for the case $\epsilon \gg 1$), in which case QND observables in that general form, with an appropriately chosen frequency operator $\hat \omega$,  can always be found. However, in the intermediate regime of the Dirac oscillator $\hat n$ is no longer a conserved quantity, a consequence of the spin-orbit induced exchange of energy between the motional and spin degrees of freedom.  

The effects of this spin-orbit coupling can be formally eliminated in the Foldy-Wouthuysen (FW) representation~\cite{moreno1989covariance}, as we now show. For the one-dimensional Dirac oscillator, the FW and the original Dirac representation are related by the unitary transformations~\cite{verlan2001excitation},
\begin{equation}
\hat{U}=
\begin{bmatrix}
\hat{A}_{n}&i\frac{\hat{B}_{n}}{\sqrt{\hat{n}}}\hat{a}^{\dagger}\\
\hat{a}\frac{\hat{B}_{n}}{\sqrt{\hat{n}}}&-i\hat{A}_{n+1}
\end{bmatrix},
\label{Umatrix}
\end{equation}
where the operators $\hat A_n$ and $\hat B_n$ are obtained by replacing $n$ by $\hat n$ in the expressions for $A_n$ and $B_n$ of Eqs.~(\ref{eq:AnBn}). The transformed Hamiltonian is then the sum of $\hat n$-conserving Hamiltonians $\hat H_{\text F}(\hat n)$ of the form
\begin{equation}
\hat H_{\text F}(\hat n)=\hat U \hat H_{\text D} \hat U^\dagger =\hat{\sigma}_{z}mc^2\sqrt{1+(2\hat{n}-\hat{\sigma}_{z}+1)\varepsilon} \, ,
\end{equation}
with eigenstates $\left |n,\uparrow\right\rangle$ and $\left |n,\downarrow\right\rangle$ and associated positive and negative eigenenergies  $E_n^+$ and $E_n^-$, respectively. Importantly,  the FW eigenstates do not involve superpositions of the two spin states, that is, the positive and negative energy eigenstates are uniquely associated with spin up and spin down, respectively, as was the case in the weakly relativistic limit $\epsilon \ll 1$. Following a similar approach we can therefore again construct a pair of QND observables
\begin{eqnarray}
\hat{X}_{1, \rm F}&=&x_{\rm zpt}(\hat{a}e^{i\hat{\omega}_\epsilon t}
                                         +{\rm h.c.})\, , \nonumber  \\
\hat{X}_{2, \rm F}&=&i x_{\rm zpt}(e^{-i\hat{\omega}_\epsilon t}\hat{a}^{\dagger}
                                         -{\rm h.c.}) \, ,
\label{eq:X12FW}
\end{eqnarray}
where the frequency operator is now 
\begin{equation}
\hat{\omega}_\epsilon\equiv \left [ \hat H_{\text F}(\hat n)-\hat H_{\text F}(\hat n-1) \right ] /\hbar,
 \label{eq:omegaFW}
\end{equation} 
and accounts for the unequally spaced energy spectrum. As in the previous cases these quadratures obey the commutation relation $[\hat X_{1,\text F},\hat X_{2, \text F}]=i\hbar/m\omega$.

From Eq.~(\ref{eq:omegaFW}), the eigenvalues $\omega_n^\pm$ of $\hat{\omega}_\epsilon$, given by $\hat\omega_\epsilon |n,\uparrow\rangle=\omega_n^+|n,\uparrow\rangle$ ($n\geq1$) and $\hat\omega_\epsilon |n,\downarrow\rangle=\omega_n^- |n,\downarrow\rangle$ ($n\geq 0$), are simply the differences in frequency of neighboring eigenstates of the FW-transformed Dirac oscillator. In the weakly relativistic limit $\epsilon \ll 1$ they approach the constant value $\omega$ characteristic of a harmonic oscillator, while for a system deep in the relativistic regime we have $\left |\omega_n^\pm\right |/\omega \to 0$ for large $n$, a signature of the increasing anharmonicity of the system. 

\begin{figure}[htbp]
\begin{center}
\includegraphics[width=0.47\textwidth]{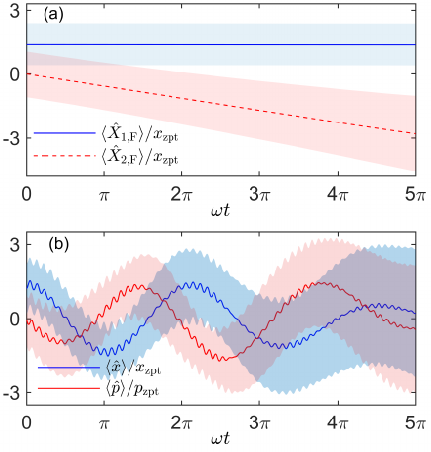}
\caption{(Color online) (a) Time evolution of the mean values (color lines) and uncertainties (background shadows) of $\hat X_{1, \text F}$ (blue) and $\hat X_{2, \text F}$ (red) under a continuous QND measurement described by the Hamiltonian~(\ref{eq:FW meas}); (b) Corresponding evolution of $\hat x$ (blue) and $\hat p$ (red). Here the Dirac oscillator is initially in a coherent state with $\alpha=0.5$, $c_1=c_2=1/\sqrt 2$, and $\epsilon=0.1$; the probe optical field is in a coherent state $|\beta=2\rangle$ and $g=0.1$. Position in units of $x_{\text {zpt}}=\sqrt{\hbar/2m\omega}$ and momentum in units of $p_{\text {zpt}}=\sqrt{m\hbar\omega/2}$.}
\label{backaction}
\end{center}
\end{figure}

If one were to ignore the important point that $\hat{X}_{1, \rm F}$ needs to be transformed back to the Dirac representation to assess possible practical measurement schemes, one might argue that it could formally be monitored e.g. by continuous QND measurements described by the total Hamiltonian
\begin{equation}
\hat{H}_{\rm{tot}}=\hat H_{\rm F}+\hbar\omega_{b}\hat{b}^\dagger\hat{b}+g\hat{X}_{1, \rm F}\hat{b}^\dagger\hat{b},
\label{eq:FW meas}
\end{equation}
where $\hat b$ represents the annihilation operator of an optical mode of frequency $\omega_b$, a form reminiscent of the optomechanical Hamiltonian of quantum optics. The associated Heisenberg equations of motion $d\hat X_{1, \rm F}/dt =0$, $d\hat X_{2,\rm F}/dt =-g\hat b^\dagger \hat b/m\omega$ and $d\hat b/dt =-i(\omega_b+g\hat X_{1, \rm F}/\hbar)\hat b$ show that the QND variable $\hat X_{1, \rm F}$ is imprinted on the phase of the optical mode $\hat b$, while $\hat X_{2, \rm F}$ is subject to the the measurement backaction from the radiation pressure force $g\hat b^\dagger \hat b$. For example, for a Dirac oscillator initially in the state $|\phi\rangle_{\rm F}=|\alpha\rangle\otimes(c_1\left|\uparrow\right\rangle+c_2 \left|\downarrow \right\rangle )$, with $|\alpha\rangle$ a coherent state and $c_1$ and $c_2$ arbitrary probability amplitudes, we have that $\Delta\hat{X}_{1, \rm F}^2(0)=\Delta\hat{X}_{2, \rm F}^2(0)=\hbar/2m\omega$. In that case, as shown in Fig.~\ref{backaction}(a), the mean value and the uncertainty in the measured observable $\hat X_{1, \rm F}$ remain constant in time as expected of QND measurements, while the mean value and variance of $\hat X_{2, \rm F}$ become$\langle\hat X_{2,\rm F}(t)\rangle =\langle\hat X_{2, \rm F}(0)\rangle-(g/m\omega)\langle \hat n_b\rangle t$ and $\Delta\hat{X}_{2, \rm F}^2 (t)=\hbar/2m\omega+(g/m\omega)^2 \Delta \hat n_b^2 t^2$.

However,  when expressed in the Dirac representation the observables $\hat X_\ell=\hat U^\dagger \hat X_{\ell, {\rm F}} \hat U$ become in general complex combinations of $\hat x$, $\hat p$, and $\bm \sigma$, so that realizing the system-probe interaction~(\ref{eq:FW meas}) is anything but obvious in general. That this is the case is also apparent from Fig.~\ref{backaction}(b), which shows the evolution of $\langle \hat x\rangle$ and $\langle \hat p\rangle$ under the measurement scheme and the same parameters as Fig~\ref{backaction}(a).  In addition to slow oscillations at a frequency of order $\omega$, the evolution of these observables is characterized by high-frequency {\it Zitterbewegung} oscillations induced by the spin-orbit coupling induced interferences between the positive and negative energy states \cite{PhysRevLett.121.110401}, as well as by significant measurement backaction. This situation is considerably more complex than for the simple harmonic oscillator, where the QND nature quadratures (\ref{eq:quadrat}) follows readily and intuitively from the simple oscillatory evolution of $\hat x$ and $\hat p$. 

\section{Implementation}

The explicit expressions of $\hat X_{1,2}$  in terms of $\hat x$, $\hat p$, and $\bm \sigma$ in the Dirac representation are in general quite cumbersome, as they include several spin-orbit-coupling composite observables (see Appendix A for their explicit form). Their measurement represents therefore a considerable challenge for arbitrary $\epsilon$, and while possible in principle, they are likely to remain unrealistic in practice. The situation is considerably simplified in both the weakly relativistic and extreme relativistic limits, as we now show. 

\subsection{Weakly relativistic limit}

For $\epsilon \ll 1$ we have
\begin{equation}
\hat X_{1,\rm nr}=\hat{x}\cos \omega t -(\hat p\hat \sigma_z/m\omega)\sin\omega t,
\end{equation}
so that in addition to the measurement of $\hat x$ the composite observable $\hat p\hat \sigma_z$ requires only spin-sensitive momentum measurements.  This is because in this limit the spin-orbit coupling is so weak as not to couple the components $\psi_1$ and $\psi_2$ of the spinor, as we have seen. $\hat \sigma_z$ is then a constant of motion and commutes with $\hat p$. Consequently, if the initial state of the Dirac oscillator is known not to exhibit spin-motion correlations, these measurements can be performed by measuring the spin and the momentum independently. 

\subsection{Extreme relativistic limit}
The situation is somewhat more complex, but still manageable in principle in the extreme relativistic limit $\epsilon \gg 1$ where
\begin{equation}
\hat{X}_{1,\rm r}=\frac{1}{2}(\hat{x}+i\hat{p}/m\omega) e^{i\hat \omega_r t}+{\rm h.c.}
\end{equation}
As we have seen, that regime is characterized by a relativistically induced anharmonicity of the oscillator which results in an excitation frequency that is no longer a constant, but rather a quantum observable $\hat \omega_{\rm r}$ that does not commute with $\hat x$ and $\hat p$, see Eq.~(\ref{eq:omegar}). In that case the measurement of the composite observables that comprise $\hat X_{1, \rm r}$ requires a coupling between the probe and $\hat x$ and $\hat p$ that is sensitive to the excitation frequency. However, as we shall see the time evolution of $\hat \omega_{\rm r}$ is far slower than the other observables, so that this coupling can be approximated by introducing an additional probe of $\hat \omega_{\rm r}$ and then combining with a probe of the motional observables. 

To see more concretely how this scheme can be implemented we consider a probe with internal degree of freedom described by the Pauli spin operators $\hat s_{ 
x,y,z}$, and external degree of freedom by the bosonic operators $\hat b$ and $\hat b^\dagger$. The measurement scheme is described by the total Hamiltonian
\begin{equation}
\hat{H}_{\rm{tot}}=\hat{H}_{\rm D}+\hbar\omega_{\rm b}\hat{b}^{\dagger}\hat{b}+\hbar\omega_{\rm{s}}\hat s_{z}+\hat V_{\rm m},\\
\end{equation}
where $\omega_b$ and $\omega_s$ are the external and the internal characteristic frequencies of the probe. in the extreme relativistic limit considered here $\hat{H}_{\rm D}\approx\hat{H}_{\rm r}$. Finally  $\hat V_{\rm{m}}$ describes the measurement interaction. Its explicit form is
\begin{equation}
\hat V_{\rm m}=\frac{\hbar\hat \omega_{\rm r}}{2} \hat s_{x}+g x_{\rm zpt} \big[\hat a(\hat s_z -i\hat s_y)+{\rm h.c.}\big ](\hat b^\dagger+\hat b),
\label{Vm}
\end{equation}
where the first term accounts for the rotation of the probe internal spin due to $\hat\omega_r$, and the second describes the Dirac oscillator driven modification of the spin-orbit coupling of the probe. The Heisenberg equations of motion ot the oscillator-probe system are
\begin{widetext}
\begin{eqnarray}
\frac{d\hat{s}_{x}}{dt}&=&-2\omega_{\rm{s}}\hat{s}_{y}-i\frac{g}{p_{\rm zpt}}[\hat{a}(\hat{s}_{z}-i\hat{s}_{y})-{\rm h.c.}](\hat{b}^{\dagger}+\hat{b}),\label{dsx}\\
\frac{d\hat{s}_{y}}{dt}&=&-\hat{\omega}_{\rm{r}}\hat{s}_{z}+2\omega_{\rm{s}}\hat{s}_{x}+\frac{g}{ p_{\rm zpt}}(\hat{a}\hat{s}_{x}+{\rm h.c.})(\hat{b}^{\dagger}+\hat{b}),\label{dsy}\\
\frac{d\hat{s}_{z}}{dt}&=&\hat{\omega}_{\rm{r}}\hat{s}_{y}+i\frac{g}{p_{\rm zpt}}(\hat{a}\hat{s}_{x}-{\rm h.c.})(\hat{b}^{\dagger}+\hat{b}),\label{dsz}\\
\frac{d\hat{\omega}_{\rm{r}}}{dt}&=&\frac{i\hat{\sigma}_{y}}{\hbar}cg(2\sqrt{\hat{n}}-\sqrt{\hat{n}+1}-\sqrt{\hat{n}-1})\hat{a}(\hat{s}_{z}-i\hat{s}_{y})(\hat{b}^{\dagger}+\hat{b})+{\rm h.c.},\label{domegar}\\
\frac{d\hat{a}}{dt}&=&-i\hat{a}{\hat{\omega}_{\rm{r}}}
-\frac{i\hat{\sigma}_{y}}{\hbar}cp_{\rm zpt}(2\sqrt{\hat{n}}-\sqrt{\hat{n}+1}-\sqrt{\hat{n}-1})\hat{a}\hat{s}_{x}
-\frac{ig}{2p_{\rm zpt}}(\hat{s}_{z}+i\hat{s}_{y})(\hat{b}^{\dagger}+\hat{b}),\label{da}\\
\frac{d\hat{b}}{dt}&=&-i\omega_b\hat{b}-i\frac{g}{2p_{\rm zpt}}[\hat a(\hat s_z-i\hat s_y)+{\rm h.c.}],
\end{eqnarray}
\end{widetext}
where the coupling between Eqs.~(\ref{domegar}) and (\ref{da}) is due to the fact $[\hat a, \hat \omega_{\text r}]\neq 0$, which leads to the difficulty of measuring $\hat x$, $\hat p$, and $\hat\omega_{\rm r}$ independently, in contrast to the weakly relativistic limit. 

Assuming that $\langle \hat a\rangle$ scales as $\sqrt{n}$,  Eq.~(\ref{domegar}) shows that $\langle\dot{\hat{\omega}}_{\rm r}\rangle$ scales then as $1/ n$, so that $\langle\dot{\hat{\omega}}_{\rm r}\rangle /\langle \hat{\omega}_{\rm r}\rangle$ scales as $g/\sqrt{n\hbar\omega m}$, where we have used Eq.~(\ref{eq:omegar}). For weak enough  measurement strength $g$, large enough oscillator excitation $ n$, and assuming further that $\langle \hat \omega_r\rangle  \gg \omega_s$, so that it dominates the evolution of the probe field, see Eqs. (\ref{dsx}), (\ref{dsy}), and (\ref{dsz}), the Heisenberg equations for $\hat s_{x,y,z}$ can be approximated as
\begin{equation}
\frac{d\hat s_x}{dt}\approx 0,\ \ \frac{d\hat s_y}{dt}\approx -\hat{\omega}_{\rm r}\hat s_z, \ \ \frac{d\hat s_z}{dt}\approx \hat{\omega}_{\rm r}\hat s_y,
\end{equation}
so that
\begin{equation}
\begin{aligned}
&\hat s_x(t)\approx\hat s_x (0),\\
&\hat s_y (t)\approx\hat s_y (0)\cos{\hat{\omega}_{\rm r}t}-\hat s_z(0)\sin{\hat{\omega}_{\rm r}t},\\
&\hat s_{z}(t)\approx\hat s_y (0)\sin{\hat{\omega}_{\rm r}t}+\hat s_z(0)\cos{\hat{\omega}_{\rm r}t}.\\
\end{aligned}
\end{equation}
Substituting these expressions into $\hat V_{\rm m}$, and for the probe initially in the internal  state $\left |\uparrow\right\rangle$ so that $\langle\hat s_{x,y}(0)\rangle=0$ and $\langle\hat s_z (0)\rangle=1$, the effective form of the measurement interaction becomes
\begin{equation}
\hat{V}_{\rm m}\approx g\hat{X}_{\rm{1,r}}(\hat{b}^{\dagger}+\hat{b}) ,
\end{equation}
so that this model detection scheme performs indeed approximate QND measurements on that observable.

\subsection{Near QND measurements as probes of relativistic effects}

The fact that in the weakly and extreme relativistic limits $\epsilon \ll 1$ and $\epsilon \gg 1$ the QND variables $\hat X_{1, \rm nr}$ and $\hat X_{1, \rm r}$ are amenable to direct detection suggests exploiting continuous measurement departures from their QND character as probes of relativistic effects, that is, of spin-orbit coupling. 

\begin{figure}[thbp]
\begin{center}
\includegraphics[width=0.48\textwidth]{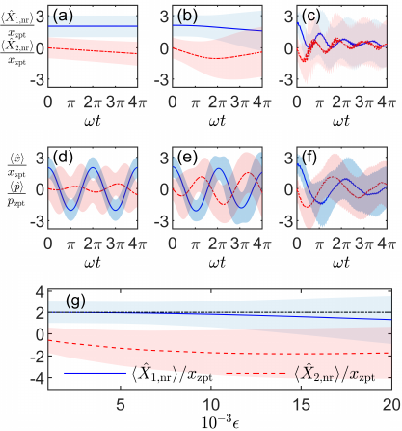}
\caption{(color online) Upper row: Time evolution of the normalized mean values (colored lines) and corresponding uncertainties (background shadows of the same colors) of $\hat{X}_{1,{\rm nr}}$ (blue solid) and $\hat{X}_{2,\rm{nr}}$ (red dashed) under continuous measurements for increasing relativistic parameters, (a) $\epsilon=0.001$, (b) $0.02$, and (c) $0.1$. Middle row: Corresponding time evolution of $\hat x$ (blue solid) and $\hat p$ (red dashed) for (d) $\epsilon=0.001$, (e) $0.02$, and (f) $0.1$. Lower row: Dependence on $\epsilon$ of $\left\langle X_{1,{\rm nr}}\right\rangle$ (blue solid) and $\left\langle X_{2,{\rm nr}}\right\rangle$ (red dashed) and their variances at time $t=4 \pi/\omega $. The Dirac oscillator is initially in a coherent state with $\alpha=1$ and $c_1=c_2=1/\sqrt 2$, the optical field is in the coherent state $\beta=1$ and the measurement strength is $g=1$. The normalized units are same as in Fig. \ref{backaction}.}
\label{backaction2}
\end{center}
\end{figure}

This is illustrated for the example of $\hat X_{1, \rm nr}$ in Fig.~\ref{backaction2}, which shows both its evolution and the evolution of its conjugate variable $\hat X_{2, \rm nr}$ for three values of $\epsilon$. In the weakly relativistic regime $\epsilon = 10^{-3}$ of Figs.~2(a) and 2(d), $\hat X_{1, \rm nr}$ behaves as expected as a QND observable, but it starts departing from this behavior as $\epsilon$ increases. This is apparent also from the temporal increase in its variance, as well as by the onset of Zitterbewegung superposed to damped oscillations at frequency $2\omega$ for large enough $\epsilon$. Fig.~2(g) shows the dependence on $\epsilon$ of $X_{1,{\rm nr}}$ and $X_{2,{\rm nr}}$ at a fixed time $t=4 \pi/\omega $, for the same parameters, illustrating the sensitivity of both $\langle \hat X_{1, \rm nr} \rangle$ and its variance on $\epsilon$. 

More quantitatively, on the  weakly relativistic side, and keeping terms up to the first order of $\epsilon$ in $\hat H_{\rm D}$, the Heisenberg equations for $\hat X_{\ell, \rm{nr}}$ become
\begin{eqnarray}
\frac{d \hat X_{1, \rm{nr}}}{dt} &\approx&-\frac{\hat \omega_{\rm{nr}}\epsilon}{2}[i \hat X_{1, \rm{nr}} + (2\hat n - \hat \sigma_z +1) \hat X_{2, \rm{nr}}],\\
\frac{d \hat X_{2, \rm{nr}}}{dt} &\approx&-\frac{\hat \omega_{\rm{nr}}\epsilon}{2}[i \hat X_{2, \rm{nr}} - (2\hat n - \hat \sigma_z +1) \hat X_{1, \rm{nr}}] - \frac{g\hat b^\dagger \hat b}{m\omega}. \nonumber
\end{eqnarray}
which illustrates the $\epsilon$-dependence of  measurement induced backaction on $\hat X_{1, \rm{nr}}$. (The results of Fig.~\ref{backaction2} were obtained numerically, keeping $\epsilon$ to all orders.) 

Similarly, slightly departing from the extreme relativistic limit the Heisenberg equations for $\hat X_{\ell, \rm{r}}$ are
\begin{eqnarray}
\frac{d \hat X_{1, \rm{r}}}{dt} &\approx&\frac{1}{8\epsilon} [ (i\hat X_{1, \rm{r}} - \hat X_{2, \rm r}) \frac{\hat{\omega}_{\rm r}}{\hat n}- {\rm h.c.}],\\
\frac{d \hat X_{2, \rm{r}}}{dt} &\approx&\frac{1}{8\epsilon} [(\hat X_{1,\rm r}+i\hat X_{2, \rm{r}}) \frac{\hat{\omega}_{\rm r}}{\hat n}+{\rm h.c.}]- \frac{g\hat b^\dagger \hat b}{m\omega}. \nonumber
\end{eqnarray}

Turning finally to order-of-magnitude estimates of relevance for a potential laboratory implementation of this system, we note first that for a sample of electrons with mass $m_e\sim 10^{-31}$kg in the relativistic limit with $\epsilon\sim 10^{3}$ and in highly excited states $n\sim 10^{4}$, the frequency of the Dirac oscillator is $\omega\sim10^{23}$Hz, and the uncertainty $\Delta X_{1}\sim10^{-13}$m, This corresponds to electron energies of the order of $E_{e}\sim\langle\hat H_{\rm r}\rangle\sim 10$MeV, rendering an experimental implementation challenging. On the other hand, these numbers are significantly improved in tabletop simulations of relativistic quantum systems. For example, for a Dirac oscillator simulated by a spin-orbit coupled ultracold atomic sample \cite{PhysRevLett.121.110401} with effective mass $m\sim 10^{-27}$kg and effective velocity of light $c\sim 10^{-2}$m/s, and for a same relativistic limit condition, the oscillating frequency $\omega\sim 10^6$Hz and $\Delta X_{1}\sim10^{-7}$m, with characteristic energy levels decreased to a manageable $10^{-9}$eV.

\section{Conclusion and outlook}

We have established that despite the existence of anharmonicity and spin-orbit coupling resulting from relativistic effects one can identify a pair of conjugate QND observables of the one-dimensional Dirac oscillator. They can be monitored relatively simply in the weakly and strongly relativistic limits of the oscillator, and may be used for instance as probes of modulated spin-orbit coupling by relativistic effect. A laboratory implementation should be possible in artificial Dirac systems. 
Although a single-particle theory, the existence of these observables may provide a simple model to investigate aspects of relativistic quantum measurements~\cite{PhysRevD.66.023510} and the generalized relativistic uncertainty principle~\cite{PhysRev.40.569,al2009minimal,todorinov2019relativistic}, and may also find applications in the development of alternative detection technique for topological insulators and other quantum systems with genuine or synthetic spin–orbit coupling \cite{RevModPhys.91.015005}.

\begin{acknowledgments}
We acknowledge enlightening discussions with Baiqiang Zhu. This work was supported by the National Natural Science Foundation of China (Grants No.~11974116 and No.~11654005), the National Key Research and Development Program of China (Grant No.~2016YFA0302001), the Fundamental Research Funds for the Central Universities, the Shanghai Municipal Science and Technology Major Project under Grant No. 2019SHZDZX01, the Chinese National Youth Talent Support Program and and the Shanghai Talent Program.
\end{acknowledgments}

\newpage

\section*{Appendices}

\subsection{Transforms between the Dirac representation and the Foldy–Wouthuysen representation}

In the Dirac representation the Hamiltonian of a (1+1)-dimensional Dirac oscillator can be written in the form of anti-Jaynes-Cummings model, 
\begin{equation}
\hat{H}_{\rm{D}}=\eta\hat{\sigma}_{-}\hat{a}+\eta^\ast\hat{\sigma}_{+}\hat{a}^\dagger+mc^2\hat{\sigma}_{z},
\end{equation}
where the matrix forms of the Pauli operators are 
\begin{equation}
\hat{\sigma}_{z}=\begin{bmatrix}1&0\\0&-1\end{bmatrix},\ 
\hat{\sigma}_{+}=\begin{bmatrix}0&1\\0&0\end{bmatrix},\ 
\hat{\sigma}_{-}=\begin{bmatrix}0&0\\1&0\end{bmatrix}.
\end{equation}
Then its eigenstates are in the form of dressed states as shown in Eq. (\ref{dressedstate}),
\begin{equation}
\begin{aligned}
&|E^{+}_{n}\rangle=A_{n}|n,\uparrow\rangle-iB_{n}|n-1,\downarrow\rangle,\\
&|E^{-}_{n}\rangle=B_{n+1}|n+1,\uparrow\rangle+iA_{n+1}|n,\downarrow\rangle.
\end{aligned}
\end{equation}

In the Foldy-Wouthuysen (FW) representation, the complications associated with the spin-orbit coupling are formally eliminated. 
The form of the Hamiltonian becomes into a diagonal $2\times2$ matrix 
\begin{equation}
\begin{aligned}
    \hat{H}_{\rm{F}}(\hat{n})&=\hat{U}\hat{H}_{\rm{D}}\hat{U}^{\dagger}\\
    &=mc^2\begin{bmatrix}\sqrt{1+2\hat{n}\epsilon}&0\\
    0&-\sqrt{1+2(\hat{n}+1)\epsilon}
    \end{bmatrix}.\\
\end{aligned}
\end{equation}
where the unitary transformation operators $\hat{U}$ is given in Eq. (\ref{Umatrix}) and we have used the equality 
\begin{equation}
    \hat{a}f(\hat{n})=f(\hat{n}+1)\hat{a}.
\end{equation}
The eigenstates become $|n,\uparrow\rangle$ and $|n,\downarrow\rangle$ corresponding to the positive eigenenergies $E^{+}_{n}$ and the negative eigenenergies $E^{-}_{n}$, respectively.
The eigenstates in the FW and the Dirac representations are related by the unitary transformation
\begin{equation}
\begin{aligned}
&|n,\uparrow\rangle=\hat{U}|E^{+}_{n}\rangle,\\
&|n,\downarrow\rangle=\hat{U}|E^{-}_{n}\rangle.
\end{aligned}
\end{equation}

Then the QND observables for a (1+1)-dimensional Dirac oscillator can be concisely expressed in the FW representation,
\begin{equation}
\begin{aligned}
&\hat{X}_{1,\rm{F}}=\sqrt{\frac{\hbar}{2m\omega}}(\hat{a}e^{i\hat{\omega}_{\epsilon}t}
                                         +e^{-i\hat{\omega}_{\epsilon}t}\hat{a}^{\dagger}),\\
&\hat{X}_{2,\rm{F}}=i\sqrt{\frac{\hbar}{2m\omega}}(e^{-i\hat{\omega}_{\epsilon}t}\hat{a}^{\dagger}
                                         -\hat{a}e^{i\hat{\omega}_{\epsilon}t}), \label{X12FW}
\end{aligned}
\end{equation}
where we defined a frequency operator  
\begin{equation}
\hbar\hat{\omega}_\epsilon=\hat H_{\rm F}(\hat n)-\hat H_{\rm F}(\hat n-1).
\label{omegaFW}
\end{equation}  
We have the corresponding minimum uncertainty state
\begin{equation}
\left | \phi \right\rangle_{\rm F}=\left | \alpha \right\rangle \otimes (c_1\left |\uparrow\right\rangle+c_2 \left |\downarrow \right\rangle).
\end{equation}

If we go back to the Dirac representation, the QND observables become
\begin{equation}
\hat X_{\ell,\rm D}=\hat{U}^{\dagger}\hat{X}_{\ell,\rm{F}}\hat{U},\, \, \ell=1,2,
\label{X12DDD}
\end{equation}
where the operator $\hat{a}$ and $\hat{a}^{\dagger}$ are transformed into the forms with complex spin-orbit coupling,
\begin{widetext}
\begin{equation}
\begin{aligned}
&\hat U^\dagger\hat{a}\hat U=\begin{bmatrix}(\hat{A}_{n}\hat{A}_{n+1}+\frac{\sqrt{\hat{n}}}{\sqrt{\hat{n}+1}}
\hat{B}_{n}\hat{B}_{n+1})\hat{a}&
i(\frac{\hat{A}_{n}\hat{B}_{n+1}}{\sqrt{\hat{n}+1}}-
\frac{\hat{A}_{n+1}\hat{B}_{n}}{\sqrt{\hat{n}}})\hat{a}^{\dagger}\hat{a}
+i\frac{\hat{A}_{n}\hat{B}_{n+1}}{\sqrt{\hat{n}+1}}
\\
i\hat{a}(\frac{\hat{A}_{n}\hat{B}_{n+1}}{\sqrt{\hat{n}+1}}-
\frac{\hat{A}_{n+1}\hat{B}_{n}}{\sqrt{\hat{n}}})\hat{a}
&(\hat{A}_{n+1}\hat{A}_{n+2}+\frac{\sqrt{\hat{n}+2}}{\sqrt{\hat{n}+1}}
\hat{B}_{n+1}\hat{B}_{n+2})\hat{a}\end{bmatrix}, \label{eq:wideeq}\\
&\hat U^\dagger\hat{a}^{\dagger}\hat U=\begin{bmatrix}\hat{a}^{\dagger}
(\hat{A}_{n}\hat{A}_{n+1}+\frac{\sqrt{\hat{n}}}{\sqrt{\hat{n}+1}}
\hat{B}_{n}\hat{B}_{n+1})&
-i\hat{a}^{\dagger}(\frac{\hat{A}_{n}\hat{B}_{n+1}}{\sqrt{\hat{n}+1}}-
\frac{\hat{A}_{n+1}\hat{B}_{n}}{\sqrt{\hat{n}}})\hat{a}^{\dagger}
\\
-i\hat{a}^{\dagger}\hat{a}(\frac{\hat{A}_{n}\hat{B}_{n+1}}{\sqrt{\hat{n}+1}}-
\frac{\hat{A}_{n+1}\hat{B}_{n}}{\sqrt{\hat{n}}})-i\frac{\hat{A}_{n}\hat{B}_{n+1}}{\sqrt{\hat{n}+1}}
&\hat{a}^{\dagger}(\hat{A}_{n+1}\hat{A}_{n+2}+\frac{\sqrt{\hat{n}+2}}{\sqrt{\hat{n}+1}}
\hat{B}_{n+1}\hat{B}_{n+2})\end{bmatrix},
\end{aligned}
\end{equation}
\end{widetext}
and the transformed frequency operator $\hat{\omega}_{\epsilon}$ is no longer diagonal in the matrix form 
\begin{equation}
\begin{aligned}
\frac{\hat U^\dagger\hat{\omega}_{\epsilon}\hat U}{\omega}=
\begin{bmatrix}\chi_{\epsilon}(\hat{n})&
i\sqrt{2\epsilon}\chi_{\epsilon}(\hat{n})\hat{a}^{\dagger} \label{omegaDD}\\
-i\hat{a}\sqrt{2\epsilon}\chi_{\epsilon}(\hat{n})&-\chi_{\epsilon}(\hat{n}+1)
\end{bmatrix},
\end{aligned}
\end{equation}
where
\begin{equation}
\chi_{\epsilon}(\hat{n})=\frac{1}{\epsilon}
(1-\sqrt{1-\frac{2\epsilon}{1+2\epsilon\hat{n}}}).
\end{equation}
The minimum uncertainty state in the Dirac representation becomes
\begin{equation}
|\phi\rangle_{\rm D}=\hat U^\dagger |\phi\rangle_{\rm F}=e^{-\frac{1}{2}|\alpha|^2}
\sum_{n=0}^{\infty}\frac{\alpha^n}{\sqrt{n!}}(c_1 |E_n^+\rangle+c_2 |E_n^-\rangle).
\label{phiDD}
\end{equation} 

\subsection{QND observables, weakly relativistic limit}
In the weakly relativistic limit $\epsilon\sim 0$, we have
\begin{equation}
\hat{A}_{n}\approx 1-\hat{n}\epsilon/4,\ \ \hat{B}_{n}\approx\sqrt{\hat{n}\epsilon}/\sqrt{2}.
\end{equation}
To the first order of $\epsilon$, the transforms (\ref{eq:wideeq}) of the operators 
$\hat{a}$ and $\hat{a}^{\dagger}$ are approximated as
\begin{equation}
\hat U^\dagger\hat{a}\hat U\approx\hat{a}-\frac{\epsilon}{4}\hat{\sigma}_{z}\hat{a}
+i\sqrt{\frac{\epsilon}{2}}\hat{\sigma}_{+},
\end{equation}
and because
\begin{equation}
\chi_{\epsilon}(\hat{n})\approx 1+(\frac{1}{2}-2\hat{n})\epsilon,
\end{equation}
the transformed frequency operator is approximated as
\begin{equation}
\frac{\hat U^\dagger\hat{\omega}_{\epsilon}\hat U}{\omega}\approx\hat{\sigma}_{z}+\epsilon
-(2\hat{n}+\frac{1}{2})\epsilon\hat{\sigma}_{z}
+i\sqrt{2\epsilon}(\hat{a}^{\dagger}\hat{\sigma}_{+}
-\hat{a}\hat{\sigma}_{-}).
\end{equation}
So in the Dirac representation for a vanishing $\epsilon$, retaining the leading order terms the QND observables (\ref{X12FW}) are still in the form of quadrature operators
\begin{eqnarray}
\hat{X}_{1}\approx\hat{X}_{1, \rm{nr}}&=&\sqrt{\frac{\hbar}{2m\omega}}(\hat{a}e^{i\omega\hat{{\sigma}}_{z}t}
+\hat{a}^{\dagger}e^{-i\omega\hat{{\sigma}}_{z}t}), \label{X12nr}\nonumber\\
&=&\hat{x}\cos{\omega t}-\frac{\hat{p}\hat{\sigma}_{z}}{m\omega}\sin{\omega t}, \nonumber\\
\hat{X}_{2}\approx\hat{X}_{2, \rm{nr}}&=&-i\sqrt{\frac{\hbar}{2m\omega}}(\hat{a}e^{i\omega\hat{{\sigma}}_{z}t}
-\hat{a}^{\dagger}e^{-i\omega\hat{{\sigma}}_{z}t}), \nonumber\\
&=&\hat{x}\hat{\sigma}_z\sin{\omega t}+\frac{\hat{p}}{m\omega}\cos{\omega t},
\end{eqnarray}
associated with a weakly relativistic minimum uncertainty state (\ref{phiDD}),
\begin{equation}
|\phi\rangle_{\rm D}\approx|\alpha\rangle\otimes(c_{1}|\uparrow\rangle+ic_{2}|\downarrow\rangle).
\end{equation}

When the Dirac oscillator departs from the weakly relativistic limit, its Hamiltonian, retaining the small terms of next order, becomes  
\begin{equation}
    \hat{H}_{\rm{D}}\approx \hat H_{\rm{nr}}-\frac{\hbar\omega}{8}\hat{\sigma}_{z}(2\hat{n}-\hat{\sigma}_{z}+1)^2\epsilon.
\label{phiHnr}
\end{equation}
To the present relativistic level, the QND observables should also include the small terms of next order,
\begin{equation}
\begin{aligned}
\hat{X}_{1}&\approx\sqrt{\frac{\hbar}{2m\omega}}(\hat{a}e^{i\omega\hat{\sigma}_{z}t}
+i\sqrt{\frac{\varepsilon}{2}}\hat{\sigma}_{+}e^{i\omega\hat{\sigma}_{z}t}
+h.c)\\
&+\sqrt{\frac{\hbar}{2m\omega}}\sqrt{2\varepsilon}[(\hat{a}^{\dagger}\hat{\sigma}_{+}
-\hat{a}\hat{\sigma}_{-})\hat{a}^{\dagger}+h.c]\sin \omega t,\\
\hat{X}_{2}&\approx -i\sqrt{\frac{\hbar}{2m\omega}}(\hat{a}e^{i\omega\hat{\sigma}_{z}t}
+i\sqrt{\frac{\varepsilon}{2}}\hat{\sigma}_{+}e^{i\omega\hat{\sigma}_{z}t}
-h.c)\\
&-i\sqrt{\frac{\hbar}{2m\omega}}\sqrt{2\varepsilon}[(\hat{a}^{\dagger}\hat{\sigma}_{+}
-\hat{a}\hat{\sigma}_{-})\hat{a}^{\dagger}-h.c]\sin \omega t,
\end{aligned}
\end{equation}
in which except the terms in Eq. (\ref{X12nr}) there are additional terms involving complex spin-orbit coupling. As the Dirac oscillator slides deeper into the relativistic regime, the QND observables have increasingly complex forms, which means their measurements are more challenging.

\subsection{QND observables, extreme relativistic limit}

In the extreme relativistic limit $\epsilon\gg1$, we have
\begin{equation}
\hat A_{n}\approx \hat B_{n}\approx1/\sqrt{2},
\end{equation}
and then the transforms on $\hat{a}$ becomes
\begin{equation}
\begin{aligned}
\hat a_{\rm r}&=\hat U^\dagger\hat{a}\hat U\\
&\approx\frac{1}{2}(1+\frac{\sqrt{\hat{n}}}{\sqrt{\hat{n}+1}})\hat{a}
+(\frac{\sqrt{\hat{n}+2}-\sqrt{\hat{n}}}{\sqrt{\hat{n}+1}})\frac{1-\hat{\sigma}_{z}}{4}\hat{a}\\
&+\frac{i}{2}[(\frac{1}{\sqrt{\hat{n}+1}}-\frac{1}{\hat{n}})\hat{a}^{\dagger}\hat{\sigma}_{+}
+\hat{\sigma}_{-}\hat{a}(\frac{1}{\sqrt{\hat{n}+1}}-\frac{1}{\hat{n}})]\hat{a}\\
&+\frac{i}{2}\frac{\hat{\sigma}_{+}}{\sqrt{\hat{n}+1}},
\label{arard}
\end{aligned}
\end{equation}
and the operator $\chi_{\epsilon}(\hat{n})$ is approximated as
\begin{equation}
\chi_{\epsilon}(\hat{n})\approx\frac{1}{\epsilon}(1-\sqrt{1-\frac{1}{\hat{n}}}),
\end{equation}
such that the transformed frequency operator $\hat{\omega}_{\epsilon}$ becomes
\begin{equation}
\begin{aligned}
\frac{\hat \omega_{\rm r}}{\omega}=\frac{\hat U^\dagger\hat{\omega}_{\epsilon}\hat U}{\omega}\approx\sqrt{\frac{2}{\epsilon}}
\begin{bmatrix}0&
i(1-\sqrt{1-\frac{1}{\hat{n}}})\hat{a}^{\dagger}\\
-i\hat{a}(1-\sqrt{1-\frac{1}{\hat{n}}})&0
\end{bmatrix}.
\end{aligned}
\label{omegarr}
\end{equation}
Then the QND observables are in the forms
\begin{equation}
\begin{aligned}
\hat{X}_{1,\rm r}&\approx\sqrt{\frac{\hbar}{2m\omega}}(\hat{a}_{\rm r} e^{i\hat\omega_{\rm r} t}
+e^{-i\hat\omega_{\rm r} t}\hat{a}_{\rm r}^{\dagger}),\\
\hat{X}_{2,\rm r}&\approx-i\sqrt{\frac{\hbar}{2m\omega}}(\hat{a}_{\rm r} e^{i\hat\omega_{\rm r} t}
-e^{-i\hat\omega_{\rm r} t}\hat{a}_{\rm r} ^{\dagger}),
\end{aligned}  
\end{equation}

In the high excited level $n\gg1$, the operators $\hat a_{\rm r}$ and $\hat a_{\rm r}^\dagger$ can be approximated as $\hat{a}$ and $\hat{a}^{\dagger}$, respectively, and then
\begin{equation}
\begin{aligned}
&\hat{X}_{1,\rm{r}}\approx\frac{1}{2}(\hat{x}+\frac{i\hat{p}}{m\omega})e^{i\hat{\omega}_{\rm{r}}t}+h.c., \label{X1r}\\
&\hat{X}_{2,\rm{r}}\approx-\frac{i}{2}((\hat{x}+\frac{i\hat{p}}{m\omega})e^{i\hat{\omega}_{\rm{r}}t}-h.c.),
\end{aligned} 
\end{equation}
where 
\begin{equation}
\hat{\omega}_{\rm r}\approx\omega\sqrt{2/\varepsilon}(\sqrt{\hat{n}-1}-\sqrt{\hat{n}})\hat{\sigma}_{y}, 
\end{equation}
associated with the approximate minimum uncertainty state 
\begin{equation}
|\phi\rangle_{\rm D}\approx |\alpha\rangle\otimes\frac{(c_{1}+c_{2})\left |\uparrow\right\rangle
-i(c_{1}-c_{2})\left |\downarrow\right\rangle}{\sqrt{2}}.
\end{equation}

\nocite{*}

\bibliography{ref}

\end{document}